\documentstyle[12pt]{article}
\input epsf
\newcommand{\be}{\begin{equation}}
\newcommand{\ee}{\end{equation}}
\newcommand{\bea}{\begin{eqnarray}}
\newcommand{\eea}{\end{eqnarray}}
\newcommand{\lb}{\label}
\begin{document}
\begin{titlepage}
\title{Classicality of primordial fluctuations and Primordial Black Holes}
\author{D. Polarski$^{1,2}$\\
\hfill \\
$^1$ LPM, UMR 5825, Univ. Montpellier II,\\ 
34095 Montpellier Cedex 05 (France)\\
\hfill \\
$^2$ LPMT, UPRES-A 6083, Univ. de Tours,\\ 
Parc de Grandmont, F-37200 Tours (France)}
 
\date{\today}

\maketitle

\begin{abstract}
The production of Primordial Black Holes (PBH) from inflationary perturbations provides a 
physical process where the effective classicality of the fluctuations does not hold for 
certain scales. For adiabatic perturbations produced during inflation, this range of scales 
corresponds to PBH with masses $M\ll 10^{15}$ g. For PBH with masses $M\sim M_H(t_e)$, 
the horizon mass at the end of inflation, the generation process during the preheating stage could 
be classical as well, in contrast to the formation of PBH on these scales by adiabatic 
inflationary perturbations.
For the non evaporated PBH, the generation process is essentially classical. 
\end{abstract}

\end{titlepage}

\section{Introduction}
We have now in cosmology a successful paradigm based on the existence of 
an inflationary stage, a stage of accelerated expansion which is sufficiently 
long to blow up the scale factor by a factor of at least 65 e-folds. During this 
inflationary stage, fluctuations of quantum origin of the scalar field(s) driving 
the inflationary phase are produced. As these fluctuations are stretched up to scales 
much larger than the Hubble radius at the time when they were produced, they can 
explain all inhomogeneities we see today even on the largest cosmological scales.
The quantum to classical transition of these fluctuations is of fundamental 
importance as the inhomogeneities we observe do not display any property typical of 
their quantum origin. As was shown earlier, this transition is actually a generic 
feature of all inflationary perturbations. The very mechanism by which inflationary 
perturbations are produced on cosmological scales leads to their effective classicality.   

On large cosmological scales probed by the observations, the quantum origin evades 
observation in the following sense: fluctuations on large cosmological scales appear to us as 
{\it random classical} quantities \cite{PS96,AFJP94}. More precisely, they have a stochastic amplitude, 
with 
a probability density deriving from the wave function of the 
fluctuations quantum state, a fixed temporal phase, and probabilities can be consistently 
assigned to classical trajectories in phase-space \cite{DP99}.
These are the characteristics of the inflationary fluctuations which allow a 
successful qualitative description of the Cosmic Microwave Background (CMB)
temperature anisotropy on all angular scales, as well as the formation of Large 
Scale Structure (LSS) through gravitational instability.

It was realized already some time ago that a spectrum of primordial 
fluctuations can lead to the formation of Primordial Black Holes and that 
these could have some observational consequences \cite{CH74,C75,NPSZ79}.    
For this aim, a spectrum of primordial fluctuations is needed. Once its 
statistics, amplitude, etc. are known, it is possible in principle to 
compute the abundance of produced PBH.
This spectrum does not have to be of quantum origin, on the contrary: 
what was considered initially was a spectrum of classical primordial 
fluctuations. Inflationary models can produce the fluctuations spectrum relevant 
for the formation of PBH.
Actually the formation of PBH is a new addition to the set of existing 
observational constraints on inflationary models \cite{CL93,CGL94,GL97,K00}, 
including pre-big bang inflationary models \cite{CLLW98} and they are in turn related to other 
cosmological observations \cite{KLM99,Y00}.
In early papers on PBH formation, the statistics of the fluctuations was taken 
for simplicity to be Gaussian and this fits well most of the inflationary 
models. However the new feature of the inflationary fluctuations comes precisely from 
what makes the inflationary paradigm successful: the generation of cosmological primordial 
fluctuations from quantum fluctuations. 

While the classical behaviour of the inflationary fluctuations is definitely very accurate 
for the description of the Cosmic Microwave Background temperature anisotropy and Large Scale 
Structure formation, we will see that in the context of PBH formation this is not always the case. 
Indeed the very mechanism by which PBH are produced does not guarantee the classicality 
of the primordial fluctuations on all scales, and for the corresponding formation times.
Actually, as we will see there is a range of PBH mass scales for which the classicality 
cannot be assumed. This of course leads to some interesting questions at the 
level of basic principles. We will first review the formation of PBH and the quantum to classical 
transition of inflationary fluctuations. This will allow us to pinpoint when and why the classicality 
of the fluctuations producing the PBH no longer holds. 

\section{PBH formation}

We consider first how PBH can be generated. It was realized long ago, before the advent of the inflationary scenario that a spectrum of primordial density fluctuations, unavoidable for galaxy formation, 
would inevitably lead to the production of primordial black holes.

For simplicity, we assume that a PBH is formed when the density contrast averaged over 
a volume of the (linear) size of the Hubble radius satisfies $\alpha \leq \delta \leq \beta$,
and that the PBH mass $M$ is just the ``horizon mass'' $M_H$, the mass contained inside the 
Hubble volume. 
%
%The mass $M$ needs not be exactly the horizon mass but it is of this order.
%
A physical scale $R(t)$ is defined by the corresponding wavenumber $k$, and it evolves with time 
according to $R(t)= 2\pi \frac{a(t)}{k}$. For a given physical scale $R(t)$, the ``horizon'' crossing 
time $t_k$ -- here we do not mean ``horizon'' crossing during inflation but after inflation -- 
is the time when that scale {\it reenters} the Hubble radius. 
Here we do not mean ``horizon'' crossing during inflation but after inflation, which will inevitably 
take place for scales that are larger than the Hubble radius at the end of inflation. 
We will simply assume that it is at that time $t_k$ that a PBH might form with mass $M_H(t_k)$. 
Obviously, there is a one to one correspondence between $R(t_k), M_H(t_k)$ and $k$. 
We can of course also make the correspondence at any other initial time $t_i$ and then relate the 
various physical quantities at both times $t_i$ and $t_k$. 
Therefore, PBH might form on scales as small as the Hubble radius right after the transition to the 
radiation dominated phase at the end of inflation. Hence the earliest PBH formation time $t_e$
corresponds to the end of the inflationary stage when the lightest PBH with $M\sim M_H(t_e)$ are 
formed.  

For concreteness, we assume further that the primordial fluctuations obey a Gaussian 
statistics with probability density $p(\delta)$, where $\delta$ is the density contrast averaged over 
a sphere of radius $R$, given by
\be
p(\delta) = \frac{1}{\sqrt{2\pi}~\sigma (R)}~ e^{-\frac{\delta^2}{2 \sigma^2(R)}}~.
\ee 
The dispersion $\sigma^2(R)\equiv \sigma^2_R = \Bigl \langle \Bigl ( \frac{\delta M}{M}  
\Bigr )_R^2 \Bigr \rangle$ is computed using a window function $W_R$
\be
\sigma^2_R = \frac{1}{2\pi^2}\int_0^{\infty}dk ~k^2 ~P(k)~W_R^2(k)\lb{sigW}~,
\ee
where $W_R(k)$ stands for the Fourier transform of the window function (usually one uses a Top Hat or 
a Gaussian window function but this is unimportant for the problem considered here) 
divided by the probed volume $V_W$, while $P(k)$ is the power spectrum of 
$\delta\equiv \delta \rho / \rho$.
Therefore the probability $P(M)$ that a PBH is formed on a scale $R(t_k)$, 
with a mass $M\equiv M_H(t_k)$ equal to the mass contained inside the Hubble radius when that scale 
reenters the Hubble radius, is given by 
\be
P(M) = \frac{1}{\sqrt{2\pi}~\sigma (R)}~ \int_{\alpha}^{\beta} ~e^{-\frac{\delta^2}{2 \sigma^2(R)}}~,
\ee 
where $M$ stands for $M_H(t_k)$, $R$ stands for $R(t_k)$ and $\sigma^2(R)\equiv\sigma_H^2(t_k)$.
Using the primordial fluctuation spectrum produced by an inflationary model, one can compute 
the production of PBH.

\section{Effective classicality of the perturbations}
The crucial point is that, independently of whether this description is acurate or not 
(we will return to that point later and consider also near critical collapse of PBH \cite{NJ98}), 
all the technical details given above fundamentally assume that the 
underlying spectrum of fluctuations is classical: the fluctuations and therefore their 
Fourier transform are all classical stochastic quantities and the power spectrum $P(k)$ can therefore 
be treated as a classical power spectrum. Though, the same expressions arise in the description of 
inflationary perturbations, the physics behind the formulas can be radically different if the classicality 
of the fluctuations is not guaranteed. As said before, this effective classicality is guaranteed for 
cosmological perturbations probed by the CMB anisotropy and LSS formation.

In inflationary models, density fluctuations as well as the CMB anisotropy arise from quantum 
fluctuations of the inflaton(s), the scalar field(s) driving inflation 
(we will take only one inflaton field for simplicity).
Let us see where the effective classicality of these fluctuations comes from.
As usual for linear equations, one works in Fourier space. 
%
%(but we will often drop the subscript, or the argument, ${\vec k}$). 
%
We will follow closely here the notation of \cite{PS96,LPS97}.
If one works in the Heisenberg representation, one makes a usual 
expansion of the (``amplitude'') operators in terms of creation and annihilation operators
\be
y({\bf k},\eta) \equiv f_k(\eta)~a({\bf k},\eta_0)+f_k^*(\eta)~a^{\dag}(-{\bf k},\eta_0)~,\lb{y} 
\ee
where $f_k$ are the field modes obeying the wave equation of the corresponding quantity.
For our purposes, it is convenient to use the equivalent expansion in terms of initial amplitudes 
and conjugate momenta
\be
y({\bf k},\eta) = \sqrt{2k}~f_{k1}(\eta)~y({\bf k},\eta_0)-\sqrt{{2\over k}}~f_{k2}(\eta)~\lb{y1}
p({\bf k},\eta_0)~,
\ee
where $f_{k1}$, resp. $f_{k2}$, corresponds to the real, resp. imaginary, part of $f_k$.

The dynamics in the inflationary case is very peculiar. Indeed, two linearly independent 
solutions can be found, here $f_{k1}$ and $f_{k2}$, with very different behaviours. 
In the course of time one of the two solutions ($f_{k1}$), typically called the ``growing'' 
mode, comes to dominate over the other one ($f_{k2}$), the ``decaying'' mode. 
At reentrance inside the Hubble radius during the radiation-dominated, or 
the matter-dominated, stage, the decaying mode is usually vanishingly small, and can therefore 
be safely neglected. A a result, the operator behaves like a stochastic classical quantity: 
the probability for a particular realization is found using an initial probability density 
distribution which evolves according to the classical equations of motion. This applies on all scales 
where the decaying mode is vanishingly small, usually the interesting ones in inflationary 
cosmology. In an alternative, but completely equivalent, language, these are the scales where the 
squeezing is tremendously high. 

Let us assume now that one would have a purely classical mechanism to produce the fluctuations 
at some primordial time $t_i$. Let us even assume that this mechanism would be able to put these 
classical perturbations on scales far beyond the Hubble radius at that time $t_i$ (which is of 
course precisely the crucial problem that is solved by inflation).
So, in a sense, for all practical purposes regarding the generation of fluctuations, 
this mechanism would do the same job as does inflation.  
Then the perturbations would have the same expansion as the inflationary ones with three notable 
differences: 
\par\noindent
a) The initial amplitudes and momenta are non-commuting operators in the inflationary case, and of course, 
by definition, classical quantities in the second case. 
\par\noindent
b) The relative amplitude of the ``decaying'' part to that of the ``growing'' part is {\it fixed} 
by quantization 
in the inflationary case and a priori free in the second case. So in order to compute the fluctuations 
on cosmological scales, one must know in the ``classical'' case, the ratio of both parts
at the time $t_i$ and specify this time $t_i$. In the inflationary case, this ratio is fixed 
initially (by quantization) at the Hubble radius crossing time during inflation, and can be computed 
at any later time.  
\par\noindent
c) Finally, even for the {\it ab initio} classical perturbations, the presence 
of the ``decaying'' part would spoil the fixed phase of the perturbations, seen for example in the 
Sakharov oscillations of baryonic fluctuations (and of course also in the acoustic peaks in the CMB 
anisotropy). In the inflationary case however, it has much more dramatic consequences: fluctuations 
are no longer undistinguishable from classical stochastic quantities. That can be seen directly from 
the fact that the Wigner function will have a width which is no longer negligible, it is negligible 
only when the decaying mode is negligible.

If the decaying part were still present, in the classical system one should use a probability density in 
phase-space in order to compute average values, etc. for the fluctuations.
Then, in the inflationary case, the quantum nature of the 
fluctuations would necessarily resurface as a probability density in phase-space does not exist.
However when the decaying part is negligible, we need only probability densities for the initial 
amplitudes, derived straightforwardly from their wave function.
This result is reinforced for a large class of environments when very high squeezing, and therefore 
vanishing of the decaying mode, holds \cite{KPS98,KP98}.
Note that for small PBH, possible mechanisms of decoherence by the environment 
(see e.g.\cite{dec,CRV01}) are not {\it a priori} restricted, like on large cosmological scales, by 
observational constraints exhibiting the fixed phase of the perturbations \cite{KLPS98}.
In inflationary perturbations vacuum states are relevant hence the 
probabilities are Gaussian, but the effective quantum to classical transition would apply to non 
vacuum initial states as well (see \cite{LPS97} for a discussion). 

Scales of interest for the generation of PBH span a range which is many orders 
of magnitude smaller than those probed by Large Scale Structure surveys and certainly than scales 
probed by CMB anisotropy observations. This implies that PBH generation gives the 
possibility to probe a very distinct part of the inflaton potential. But even more, and this is 
what we want to stress in this letter, the problem of the classicality of the primordial inflationary 
fluctuations could resurface during PBH generation.    

\section{Classicality of PBH production}
Where is now the problem with PBH? It lies in the very fact that PBH can be produced as soon as the 
fluctuations reenter the Hubble radius after inflation.
However, for the smallest possible scales, corresponding to the smallest masses, the decaying mode 
still ``had no time'' to disappear!
As a consequence, all that was said above does not apply anymore to these small scales. In particular, 
one cannot speak 
anymore about classical fluctuations with consistently assigned probabilities on which the above 
formulas rely.
In order to make quantitative estimates, let us consider the peculiar gravitational potential 
$\Phi(\bf k)$. It is simply related to the power spectrum $P(k)$ by Poisson's equation.
Its expansion is analogous to (\ref{y}) (or (\ref{y1})) \cite{LPS97}
\be
\Phi({\bf k},\eta) = \Phi_k(\eta)~a({\bf k},\eta_0)+\Phi_k^*(\eta)~a^{\dag}(-{\bf k},\eta_0)~.\lb{Phi}
\ee
Further, everything that was said about eq.(\ref{y},\ref{y1}) will apply to eq.(\ref{Phi}) for the 
quantity $\Phi$. 
Clearly, the degree to which the effective quantum to classical transition will occur 
(and therefore quantum interference will be essentially suppressed) is given by the ratio 
\be
D_k\equiv \frac{\Phi_{k,gr}}{\Phi_{k,dec}}
\ee
of the growing to the decaying mode of the peculiar gravitational potential $\Phi(\bf k)$. 
Very large values for $D$ will correspond to an effective quantum-to-classical transition.
It is physically appealing to give the result as a function of the PBH masses $M$.
Then, for adiabatic fluctuations produced during the inflationary stage, this is very elegantly 
expressed as 
\bea 
D(M) &=& 4A~ GH_k^2 ~\frac{M^2}{M^2_p}\lb{DM} \\
&\simeq& 4~ GH_k^2 ~\frac{M_H(t_e)^2}{M^2_p} ~~~~~~~~~~~{\rm for}~k\sim k_e\equiv (aH)_{t_e}~,\lb{DMe}
\eea
where $A$ is the growth factor of the peculiar gravitational potential $\Phi(\bf k)$ between the 
inflationary stage and the radiation dominated stage, while $H_k$, resp. $V_k$, is the Hubble parameter, 
resp. the inflaton potential, at Hubble radius crossing during the {\it inflationary} stage, and finally 
$M_p$ is the Planck mass.
The growth factor is model dependent, it is generically larger than one, and of order one for scales 
crossing the Hubble radius towards the end of inflation. 
The ratio $D(M)$ will grow with increasing PBH masses $M$, due essentially to the last term in 
expression (\ref{DM}). Clearly, there is a range of scales, those exiting the Hubble radius towards the 
end of inflation $t_e$, where $D$ will not be large and the quantum nature of the fluctuations is 
important. 
This effect will be maximal for masses $M\sim 1$g corresponding to the horizon mass $M_H(t_e)$ at the end 
of inflation. Equations (\ref{DM},\ref{DMe}) refer to adiabatic perturbations produced during inflation. 
It is interesting to note that precisely for those scales $k\sim k_e$, PBH could be produced by parametric 
amplification of the fluctuations, the details being model dependent \cite{BT00}. 
By a mechanism analogous to the one described in the previous Section, namely the disappearance of 
an exponenetially decaying mode, the growing mode of the parametrically amplified 
fluctuations would quickly become classical. Hence, depending on the efficiency of preheating, 
the generation process of part of the PBH formed on these very small scales could be essentially 
classical as well.  

There is a natural cut-off in the PBH masses due to PBH evaporation and it is interesting therefore 
to relate our result to this phenomenon. Indeed, it is well known that the lifetime of a Black 
Hole is simply related to its initial mass. Hence, PBH with masses less than 
$M_{*}\approx 10^{15}~$g will have either completely evaporated or in any case be in the latest 
stage of their evaporation. For these PBH, quantum mechanics resurface crucially in their 
evaporation process so that the quantum origin of the fluctuations leading to their generation becomes 
less important. And anyway, these PBH will have disappeared under the form of a radiation with a 
thermal spectrum. Hence, it is interesting to investigate in particular what happens to those surviving 
PBH, those with masses $M>M_*$. 
If we consider for example chaotic single-field slow-roll inflation 
with inflaton potential $V=\frac{1}{2}m^2\phi^2$, then the following result is obtained 
\be
D(M_{*}) \simeq 32A~ \frac{V_k}{M^4_p} ~\frac{M_*^2}{M^2_p} \sim 10^{28}\lb{D}~.
\ee
Generally, if one considers conservative values pertaining to inflationary stages, namely 
$G H_k^2 > 10^{-20}$, we still obtain a very high upper bound
\be
D(M > M_{*}) > 10^{20}~.\lb{D1}
\ee
Equations (\ref{D},\ref{D1}) have been derived assuming that the PBH mass 
is related to the formation time (and scale) as $M=M_H(t_k)$. 
Recently, it has been shown numerically that PBH formation may occur by near critical 
collapse \cite{NJ98}. 
This changes the exact relation between the PBH mass and the ``horizon'' mass $M_H(t_k)$ and 
the abundance of produced PBH as a function of mass $M$ \cite{LG99}.
In this case, eq.(\ref{DM}) will be simply replaced by  
\be
D(M) = 4A~ GH_k^2~ \frac{M_H^2}{M^2_p}~,\lb{RC}
\ee
where the PBH mass $M$ is now some function of $M_H$ which is not fixed by $M_H$ alone. 
At a given time, PBH of different masses can be produced by near critical collapse.
But it introduces no fundamental change concerning the quantum to classical transition of non 
evaporated PBH as the produced PBH at some time $t_k$ by near critical collapse cannot have masses 
much larger than $M_H(t_k)$.

As is obvious from eq.(\ref{D1}), one can safely use the effective classicality of the fluctuations 
for the non-evaporated PBH. Hence, for all PBH produced after approximately $10^{-23}$ seconds, 
quantum-to-classical transition is already extremely effective. This means that quantum interference 
for these PBH is essentially suppressed and one can really work to tremendously high accuracy with 
classical probability distributions: probability can be assigned consistently to the fluctuations 
leading to the formation of PBH not yet evaporated.

There has been some suggestions recently that transplanckian 
frequencies obey a different dispersion law and may change some of the standard results pertaining to 
the inflationary primordial fluctuations (for a review see e.g.\cite{J00}). 
Typically in inflationary scenarios, the crucial time is 
the Hubble radius crossing time. At that time the Hubble radius is much lower than the Planck mass 
$M_p$ and the usual dispersion law must hold. Put aside the question whether the claim 
based on some new transplanckian physics is consistent with observations (see \cite{S01} for a recent 
discussion), the vanishing of the decaying mode is solely 
based on subplanckian dynamics and the results derived here would be unaffected by any 
transplanckian new physics.     
In conclusion, all the scales for which the quantum nature of the fluctuations generating PBH 
could reveal itself are ``hidden'' deeply inside the scales affected by PBH evaporation. 
It is unclear whether PBH relics, if they exist at all, could unveil the quantum origin of 
the fluctuations generating them.

\end{document}